# Survey of Human Factors in Crisis Responsive Software Development

*Sergey Viktorovich Zykov*
*PhD, Professor of National Research*
*University Higher School of Economics*
*Moscow, Russia 125319*
*szykov@hse.ru*

*Joseph Afriyie Attakorah*
*M.Sc., National Research University*
*Higher School of Economics*
*Moscow, Russia 125319*
*dattakorakh@edu.hse.ru*

*Abstract*— Software development, despite all the significant improvements it contributes to society, is a very expensive high-risk venture. Every software project commences with the intention to deliver a software product on time and within budget, but a great percentage of these software projects either fail to achieve their full potential or fail entirely. A lot of factors come to play in this regard and the dominant amongst them, which is often overlooked and oftentimes the biggest or strongest factor, is the human factor, as software development is a basic human endeavor. Knowledge and understanding of human factors can greatly influence the success of software development. The purpose of this survey is to find out and throw more light on the human factors that have influences in crisis responsive software development, spell out some recommendations, and suggest future research directions. In this paper, we use a case study to investigate the human factors.

*Keywords— crisis, ergonomics, software development*

I. INTRODUCTION

The catastrophic failure of software development that leads to the incomplete and degrading performance of software products, software crisis, has haunted the field of software development for a long time. But while this requires great attention, very minimal regard is given to human factors, amongst all other factors. This issue is concluded in Laleh Pirzadeh's master thesis paper [1] that in spite of the impact of human factors on software development and the level of success/failure, there has not been enough research focused on the area. Human factors have been overlooked in software development process as well as software engineering management.

"Great software doesn't come from tools, it comes from people", by Kling [2]. Software development is a human endeavor where an individual or a group of individuals in an organization collaborate to deliver a software product. In such a situation, a variety of human factors come to play in determining the success of running such a risky collaborative venture [3]. These factors have varying levels of influence and may either be positive or negative. Software developers and stakeholders can inadvertently jeopardize software development in many ways allowing minor problems to escalate into major incidence that can only be resolved at great effort and expense.

Capretz [4] states that software product is the result of human activities. During software development, there are a lot of interactions between the parties involved; the customer/client, the project manager, and the developer/team. These interactions, intuitively, may include the management of teams, thus managerial aspect, communication between developers in the team and/or between the team and the customer, teamwork, negotiations, team composition, standards of maturity and the expressions of developer experience. These influencing factors may navigate the development of the software towards success or failure and so require very good attention.

In a crisis responsive software development, such as agile, how a developer may accept or refuse an incoming task regarding the current situation before the subsequent iteration or sprint starts as well as when and how this is communicated is a very important factor.

II. LITERATURE REVIEW

Crisis has devastated many software projects over the years, some resulting in total catastrophic failure and others acquiring minimal impact. This usually happens when it does not meet the customer requirements, when projects run over budget, run over time, when the quality of the software is low or the software product is inefficient, and when the projects is unmanageable and difficult to maintain. Crisis are caused largely by human factors. Pirzadeh's [1] mentions in his master thesis paper that as transferring from pre-planned to agile development process, a crisis responsive software development, there will be more "Interpersonal" level of human factors involved that are barely addressed by researchers so far. In this section, I present a review of related works on human factors that may influence or contribute to crisis in software development. Human factors are basically the human behaviour and interactions that are exhibited throughout the activities of software development.

A review of the managerial issues in software development by Taveras [5] informs that software development is mainly the results of human activities and as such incorporates our problem-solving skills, cognitive limitation and social interactions so it is important to understand human factors in order to administer their impact in software development. The author concludes that software development, as complex as it is, is subject to the vulnerabilities of the human cognitive and personal behaviour. Hence, software crisis could likely be avoided when human factors such as soft skills, team building, communication and management abilities are well considered. Human factors can be seen in three categories: individual, interpersonal, and organizational.

The individual category covers individual human issues related to software development. This aspect includes such factors such as individual characteristics, personalities, human psychology, cultures, decision making, managers' management skills (individuals), Personal Software Process (PSP), and individual learning and improvement [1] . The Personal Software Process (PSP) is a framework that is designed to help software engineers better bring discipline to the way they develop software [6].

The interpersonal issues refer to issues among individuals affecting or being affected by the software development process. The Interpersonal level is more concerned with the communication, collaboration and cooperation among individuals in a development team and how they affect the overall success/failure or productivity of the software development process. Organizational issues pertain to human factors and issues regarding the organization and the working environment. The human factors from organizational point of view involve management and decision making in organizations, team building, organization environment (multi-culture), industrial issues, etc. Pirzadeh [1], investigates, identifies, and characterizes human factors and their impact on software development from different perspectives some of which will be considered in this paper.

Muthengi [3] references Pressman 2010 that customer collaboration and communication is important in human factor software development, agile development context, just as developer coordination is vital in a globally distributed software development environment. In team-based software developments, conflicts are inevitable. These conflicts simply refer to any differences of opinions and misinterpretations of information such as customer requirements or Unified Modelling Language (UML). Such situations require an effective conflict resolution process under the leadership of a capable project leader/manager. The paper also sheds light on the organizational issues in software development where there might be communication and skill set balance management regarding team composition. It is important to place the right developer into the right team and assign them the right role according to their specialty. This relates to a managerial aspect where the right decision about placements of skills set can optimally enhance the success of software development. The team composition is indeed very vital. Members of the team need to be motivated in carrying out the tasks. Muthengi discusses three major factors in his paper that affect the success failure of software development, and these include the role of management, team composition, and communication. This is demonstrated in figure 1.

Viller, Bowers and Rodden [7] reviews human factors in software requirement engineering in terms of individual, group and organizational. It uses phrases such as Leadership experience, group polarization and group thinking, and managerial aspects, to describe human factors. In terms of leadership, it talks about how the level of experience in management results in making a certain vital decision that influences how software projects are run to succeed or fail. With an experienced manager, better decisions are made to keep software development and its management on track. It is vital to have at least one experienced person to help manage the development process.

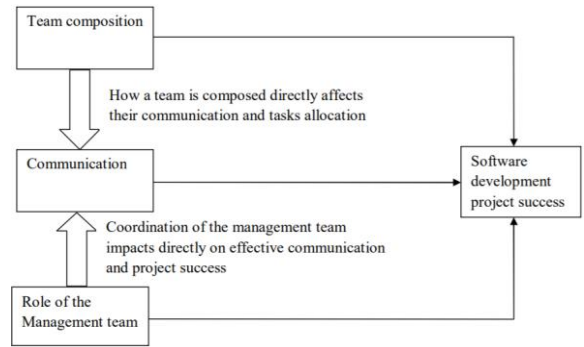

Figure 1. Teamwork, communication and management on project success [3].

Fernández, Sanz and Misra [8] analyze the contribution in this area of research and confirms that the field, as important as it is, has attracted little attention. The authors mention that adding more people to a delayed project does not necessarily lead it off failure but increases the cost. The qualification and experience of the development personnel is rather an influential aspect/human-factor. In addition, the morale and motivation of individual developers and team is recognized as a major influence. The education and abilities of the developers stand as a very important factor in the development process as it is more about people working together than it is about the tools and the defined process. This paper primarily highlights the experience level of the individual developers and manager of a project as a dramatic influential factor in the software development process.

From the above reviews, we came up with these curated human factors: communication, experience (individual/team), team composition, motivation, teamwork/collaboration, management (including soft skills), conflict resolution and problem-solving skills. We can compare these human factors in the table below:

*1) Table 1. Human factors*

| Human Factor | Description |
|---|---|
| Communication | This encompasses all the ways of sharing information and exchanging knowledge. How and when this is done has a significant impact on the development process. |
| Experience | This is how much software development a developer or the team may have done before. A rich experience can help lead the way to success while a poor experience could guide the development process off towards the wrong path. |
| Teamwork | This refers to the effective and efficient combined actions of the members involved in the development process. |

| | How well each person plays his role in the team counts a lot. |
|---|---|
| Conflict resolution | This is a process where two or more disagreeing members find a peaceful solution to their dispute. This is very important in situations when there is a misinterpretation of language (eg. UML) or there has been some sort of communication barrier. |
| Problem-solving | The process of finding solution to difficult or complex issues. Different problems frequently and occasionally occur when development is in progress so the skill of problem-solving can be very handy here. |
| Team composition | This refers to the members of the team. The specific constituent members of the team cannot be overemphasized. The members must be meticulously selected for the available roles for an optimal result. |
| Motivation | Motivation is what inspires the zeal in the team members towards the project objectives. It enables the morale in their performance. |
| Managerial Aspect | This involves the coordination of the team. All necessary managerial skills must be present to make the best decisions towards the success of the project. It is one of the most vital aspect of human factors. |

In the next section, a survey is conducted to find out about how influential the above-listed factors are in crisis responsive software development.

III. SURVEY

*A. Method*

Case Method Research

This survey uses-case study strategy as an empirical inquiry to investigate the human factors within a real-life context. A research case explores cause and effects [9].

*B. Case Study*

Following Grandon Gill's approach where he always had his best luck, our case source is acquired through networking, specifically, a professional organization. We will consider one software company in Ghana, West Africa, which we interacted with, a start-up company that was incorporated in 2016. For the sake of confidentiality, we will be close-mouthed about the name of this firm, our way of disguising the case as recommended by Grandon Gill. This software company develops and delivers web and mobile applications to clients. There were two cases of software crisis in the year 2018 one of which will be discussed here. One project had a two-member team and in the other project, a four-member team, besides the clients. We will find out the most influential human factors involved in the latter case.

*1) Table 2. Software Crisis.*

| Software | Failure Description | Casualties |
|---|---|---|
| Request and complaint ERP software prototype | Failure to include enough number of developers, from the cradle, on the team, resulted in the project being difficult to maintain, running over-time and consequently going over budget. | Several months of hard work, delayed developers' salaries and huge loss of capital |
| Website with Blog | Big gap between customer requirement and the software developed. The project run over-time and was never delivered | More than a month of hard work, a Huge loss of capital and an unsatisfied client |

*2) CASE: Developing a request and complaint web app prototype for a financial institution.*

What was meant to be the company's best shot to corner the market turned out to be catastrophic. This project was started off by a one-member team with a facile generalization of requirements and later when the deadline was closing in, handed over to a four-member team. The requirements and the existing codes were discussed and analyzed by the new team over a one-hour meeting. Because of how time was already a little far spent, and the presumed level of experience of the developer team, some details were not brought to light by the first developer. The code written by the first developer was unmanageable and difficult to develop on top of or maintain. As it was in the utmost interest of this software firm to win this contract with a clean prototype, the new team decided to rewrite the entire software from scratch with more recent development tools and frameworks. The cost of development went up and the project run over-time. Eventually, this undertaking was abandoned by the client. Several months of hard work resulted in delayed developers' salaries and a huge loss of capital and an opportunity.

*3) Analysis and Results:*

Questionnaires were administered with the team of four members on this project to find out which human factors may have contributed or influenced the most in sealing the fate of the project. The questionnaire covered these human factors: Communication, Experience, Teamwork, Conflict resolution, Problem-solving skills, Team composition, Motivation, Managerial Aspects. The respondents answered that the human factors that had the most influence on the failure of the project were communication and managerial aspect, besides all other human factors that had varying levels of effects. The feedback:

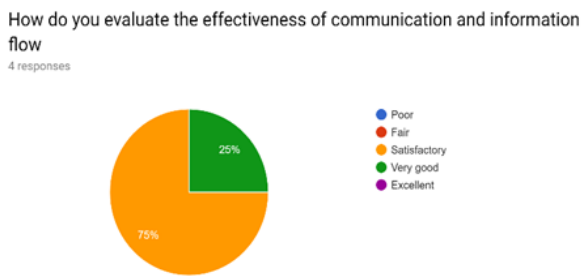

Figure 2. Response for information flow. Three respondents rated communication to be satisfactory and one rated "very good" for communication

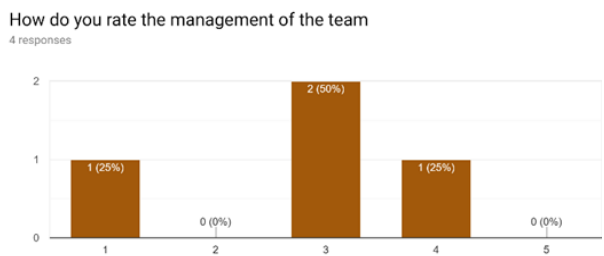

Figure 3. Response for rating management. Two respondents rated 3 (satisfactory) for managerial aspect and the other two rated 1 (poor) and 4 (very good)

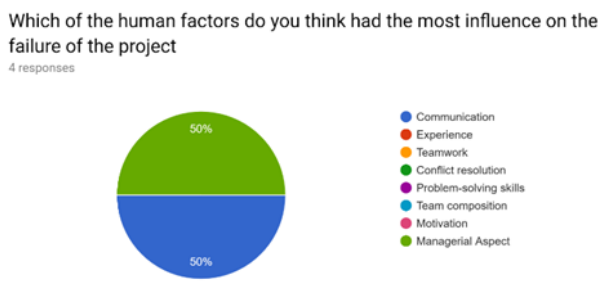

Figure 4. Most influential factor.

The response from this case study tells us that all the aforementioned human factors have significant influence on crisis responsive software development, but the managerial aspect and communication stand as the most influential human factors. This makes sense as by looking at the case we realize that it had a pretty bad start, probably with poor planning. And this has to do with the managerial aspect mostly. Communication was also a major factor given the fact that not the entire information pertaining to the project was shared with the second team while during the short meeting held.

It is recommended that these two human factors are critically considered and well enhanced/enabled for every software project at hand to reduce the chances of falling into crisis.

*C. Recommendation*

Guides such as the Team Software Process (TSP) and the Personal Software Process (PSP) are very handy means of dealing with some human factors in crisis responsive software developments.

The TSP guides developer teams in software development and helps to improve the quality and productivity of developer teams while helping them to meet cost and schedule commitments [10]. It is designed for both small and large teams.

The PSP guides individual developers in software development with a set of methods that show software engineers how to plan, measure, and manage their activities. PSP is effective for achieving software product goals within schedule and budget. It is designed for use with any programming language or design methodology and it can be used for most aspects of software work, including writing requirements, running tests, and fixing bugs [6].

The Microsoft Solution Framework (MSF), also worth mentioning, defines a set of principles, disciplines, and guidelines for delivering software services from Microsoft. This approach is not limited to software development but can also be used in a wide range of activities in technology projects, such as Developer Operations (DevOps), networking or infrastructure projects [11], [12].

There may be other means of dealing with human factors in software development, but the above-mentioned ones are very helpful to consider in human factor management.

One more thing to touch on is knowledge transfer. Since software development is a human endeavour, knowledge sharing, and communication are inevitable. Knowledge transfer is a practical problem of transferring knowledge with the objective to organize, distribute knowledge, and ensure its availability for future users. In crisis responsive software development, several factors, such as cultural issues, maturity level and mentality may enhance or obstruct knowledge. Successful knowledge transfer requires special training of the receiving side, which involves several human-related factors, such as prior knowledge, feedback, and mastery [13]. The developers' prior knowledge can help or hinder the acquisition of new knowledge and influence his interpretation of information.

"If students' prior knowledge is robust and accurate and activated at the appropriate time, it provides a strong foundation for building new knowledge. However, when knowledge is inert, insufficient for the task, activated inappropriately, or inaccurate, it can interfere with or impede new learning" [14].

In the lights of knowledge sharing and as far as human activities are concerned, feedback cannot be overemphasized. In crisis responsive software development such as Agile methodology, feedback loops are the driving factors that enable the agile team to adapt the process quickly when need be. It helps to communicate about some aspect(s) of developers' performance relative to specific target criteria, provides information to help developer progress in meeting those criteria, and is given at a time and frequency that allows it to be useful. "Goal-directed practice coupled with targeted feedback enhances the quality of students' learning" [14]. "To ensure the learning quality, adequate bidirectional feedback-driven meta-cognitive cycle organization is very important" [13]. Feedback is always important.

Finally, the climate of the development process and organizational environment may also impact knowledge transfer. "A negative climate may impede learning and performance, but a positive climate can energize students' learning" [14].

## IV. Conclusion and Future Works

Software development is a complex process that involves human activity. It is vital to not overlook but consider the quality presence of these important human factors such as communication, teamwork, experience, motivation, problem-solving skills, managerial aspect and to mention but a few. Software development teams are more likely to succeed when appropriate attention is given to these factors. Hence it is recommended that developers should consider personal development regarding these factors while the Human Resource department also considers including them in their recruitment and training processes. Enforcing the Personal Software Process (PSP) is also a very helpful choice.

An upcoming new wave of the crisis is to deal with digital technologies, AI, and Industry 4.0. The interconnection of systems with Digital technologies and Artificial Intelligence (AI) paved the way for industry 4.0. To build and manage a sustainable, scalable enterprise in today's business environment, you need to use tools that enable you to streamline tasks, boost productivity and collaboration, and leverage data in real-time. Industry 4.0 is a new phase in the digital revolution that deals with interconnectivity and Internet of Things (IoT). [15].

For future research, it would be salient to look out for crisis in industry 4.0 and the human factors that may exist in this field, as well as human factors in AI development. Further research may find out about the replacement of employees with robots and whether these replacements eliminate possible crises or improve on human factors with AI; finding out if AI is useless without human beings or vice versa is worth researching. More research done in the directions mention can bring to light all important factors to consider avoiding or managing crises.